# How Astronomers View Education and Public Outreach: An Exploratory Study

Lisa Dang[1, 2] & Pedro Russo[2]

1 McGill University, Canada; E-mail: kha.dang@mail.mcgill.ca
2 Leiden Observatory, Leiden University, the Netherlands; E-mail: russo@strw.leidenuniv.nl

## Abstract

Over the past few years, there have been a few studies on the development of an interest in science and scientists' views on public outreach. Yet, to date, there has been no global study regarding astronomers' views on these matters. Through the completion of our survey by 155 professional astronomers online and in person during the 28[th] International Astronomical Union General Assembly in 2012, we explored their development of and an interest for astronomy and their views on time constraints and budget restriction regarding public outreach activities. We find that astronomers develop an interest in astronomy between the ages of 4-6 but that the decision to undertake a career in astronomy often comes during late adolescence. We also discuss the claim that education and public outreach is regarded an optional task rather than a scientist's duty. Our study revealed that many astronomers think there should be a larger percentage of their research that should be invested into outreach activities, calling for a change in grant policies.

## 1.  Introduction

In 2004, the European Union issued a major report addressing the need for more scientists and therefore students to be involved in advanced studies of science and technology in order to achieve desired economic growth (European Commission, 2004). The European Union isn't the only one concerned about the decrease in scientists and the associated effect on the economy. In 2005, the National Academies in the United States published a report discussing the condition of science and technology which emphasized the importance of science education to the economy. During a speech following this publication, in 2006, the then President of the United States of America, George W. Bush, brought up the concern that corporations have about the imminent retirement of baby-boomer scientists over the next decade (Bush, 2006). Nowadays, the importance of science education and communication is recognized globally.

An approach to improve science education and communication, as well as developing methods to ignite interest or awareness among the public, is to understand scientists' perception of education and public outreach (EPO). In order to implement science education and communication  initiatives it is useful to identify what sparked their interest for science. Many studies have reported that the interest for science is more likely to develop at primary school but the decision to become a scientist often comes during the end of adolescence (Maltese & Tai, 2010).

Maltese and Tai (2010) investigated the importance of an early interest in science in developing scientists. Their survey received 116 responses from both practicing and retired physicists and chemists as well as graduate students across the USA. The participants were asked about the moment when they were first interested in science and what initiated this interest. This survey concludes that for the majority (about 65% of the sample), an interest in science was developed before or during primary school, which is 6 times greater than the proportion who claim to have developed an interest for science in college.

Additionally, in 2004, the Royal Society in England conducted an online survey with scientists and engineers where participants were asked when they first considered a prospective career in science (Royal Society, 2004). The majority (63%) of the respondents claimed they had thought about becoming a scientist or an engineer by the age of 14. This is another piece of evidence that shows the importance of education and public outreach and initiating interest for science at a young age.

A study conducted by Cleaves (2005) in the UK examined the factors leading students to pursue academic studies in science during college. A total of 69 students from six different schools were interviewed 4 times during a period of 2 years before entering college. The goal was to follow their thought process regarding the decisions they took concerning their career choice. Surprisingly, many of the students who chose to continue studies in science, technology, engineering and mathematics (STEM) did not enjoy high school science. Their choices were mainly based on the idea of the career they wanted and the flexibility that science studies offer. On the other hand, students who did not plan on pursuing STEM studies experienced the same lack of interest for high school science, but their exposure to science was unenjoyable enough to impede them from undertaking science studies in college.

A study conducted by Lindahl (2007) investigating factors influencing the persistence in advanced studies of science show that students claim interest in science to be one of the most important factors in their decision to pursue a career in science. The survey was conducted with 70 Swedish students between the ages of 12 to 16, and consisted of a combination of interviews and questionnaires. It revealed that these students started considering a potential scientific career as early as the age of 12. Students also reported that the way science is taught in school was usually not representative of the natural world they experience on a daily basis and did not make them more engaged in classrooms. Consequently, these experiences often affected their decisions to continue to study in science. This stresses the importance of igniting interest towards science amongst children and younger teenagers both inside and outside the classroom.

Nowadays, most scientists have a positive attitude towards science communication and education and public outreach (Andrews, 2005; Poliakoff, 2007; Ecklund, 2012), however, there is still place for improvement in the number of scientists taking part in EPO and the amount of effort that they dedicate to these activities. Many studies tried to investigate factors which motivate and inhibit scientists to undertake EPO initiatives. Ecklund's studies explored physicists and biologists' views on science outreach and revealed that many of the scientists who took part in the research claimed that one of the most common impendent factors to public engagement are time constraints. Many invest a lot of time in either research or teaching leaving very little time for EPO. Other important inhibiting factors are disapproval by mentors and department heads and the lack of career recognition from taking part in EPO activities.

Moreover, the report published by the Wellcome Trust in 2000 on the role of scientists in society revealed that many scientists (23% of the participants) think that time constraints play a significant role in preventing participating in EPO activities. 60% agreed that what was required from them each day left them very little time for education and public outreach initiatives. They also think that EPO activities are not financially advantageous. Given this general outlook of EPO initiatives, it can be rather difficult for scientists to meet the criteria science communication guidelines suggest for effective outreach projects. Many of them agree that money and time constraints often discourage from taking part in EPO activities.

While there are several studies regarding the development of interest in science at an early age, its importance in developing scientists (Maltese & Tai, 2010; Cleaves, 2005; Lindahl, 2007; The Royal Society, 2004) and factors motivating or inhibiting EPO initiatives (Poliakoff, 2007; Ecklund, 2012), there is currently no global study addressing the role of early interest in developing astronomers and their views on education and public outreach. Most studies include data from former, current and future scientists from different fields, but we would like to examine astronomers' opinion and see if it differs from one field of science to another. Moreover, little published research has studied the amount of time, effort and money scientists invest in education and public outreach. We would now like to answer the following questions:

(1) At which point in their life have astronomers developed an interest for astronomy?
(2) Are education and public outreach activities viewed as a hobby rather than a duty?
(3) How important are EPO activities according to astronomers?
(4) What do astronomers think of the budget allocated into EPO activities?

There results will then be compared to other literature and we will then see if astronomers' views differ from other fields of science. This will also serve to target when and how an interest for astronomy is initiated.

## 2. Methods

The data collected for this analysis were both quantitative and qualitative answers from a survey. A questionnaire was designed to address the development of astronomers' first interest in astronomy and their views on education and public outreach. The target groups of participants for this study were future, current, and retired professional astronomers: more specifically, astronomy students (involved in either undergraduate or graduate studies), PhD candidates, post-docs, faculty members and directors.

The International Astronomical Union, the largest association of astronomers across the globe, constitutes 11 319 individual members. In August 2012, more than 3000 astronomers gathered at the 28th General Assembly in Beijing, China, to discuss, share, present and debate the most exciting discoveries about the Universe. Universe Awareness (UNAWE), in partnership with the IAU Office of Astronomy for Development, performed individual interviews with a 61 astronomers randomly selected (and attending the IAU GA) in order to investigate when they first became interested in their field and their views on education and public outreach. Additionally, other potential participants were solicited by emails through Canadian Astronomical Society (CASCA) membership. A total of 94 responses were obtained from the online survey between December 11, 2012 and January 24, 2013.

In total, 155 responses were obtained for this study. Although these methods of sampling are, most of the time, the best option when searching for a representative sample, it restricts the generalizability of the results as the solicited potential candidates were offered the choice of not participating to the survey.

The answers provided both from the online survey and collected at the general assembly of the IAU were first combined, then preliminary descriptive statistical analysis were performed on each set of data to have an overview of the distribution. Further statistical tests were applied to confirm or disprove correlation between different variables and meaningful differences between groups of astronomers.

*Sample*

The sample consisted of 155 astronomers from 31 different countries across the globe. This included 102 males, 51 females and 2 people who did not disclose their gender. The age of the participants vary from 23 to 72, where 55% of them are within the age range of 25 to 45. The sample contains both students and professional astronomers. The majority of the sample are currently practicing astronomers including 58 faculty members, 28 post-doctoral fellows and 29 PhD candidates.

## 3.    Developing an Interest in Astronomy

The questionnaire revealed that 65% of the respondents had developed an interest for astronomy between the ages of 4 to 12. Nearly 50% of all the participants first developed an interest between 4 to 9 years old. These results are in agreement with other studies such as Lindahl (2007), Cleaves (2005), the Royal Society (2004) Maltese & Tai (2010), which reported that scientists from other fields interest in science first occurred in during elementary school and early adolescence as well.

As the data for the age at which astronomers' were first excited about astronomy is not normally distributed as shown in Figure 1, the Maan-Whitney's U test, a non-parametric test, was used to examine the gender difference in interest development. The analysis showed that there is no significant difference in the distribution of the age of interest development within astronomers ($p > 0.05$). This was expected as it agrees with the Maltese & Tai (2010) studies on sources of early interest in science.

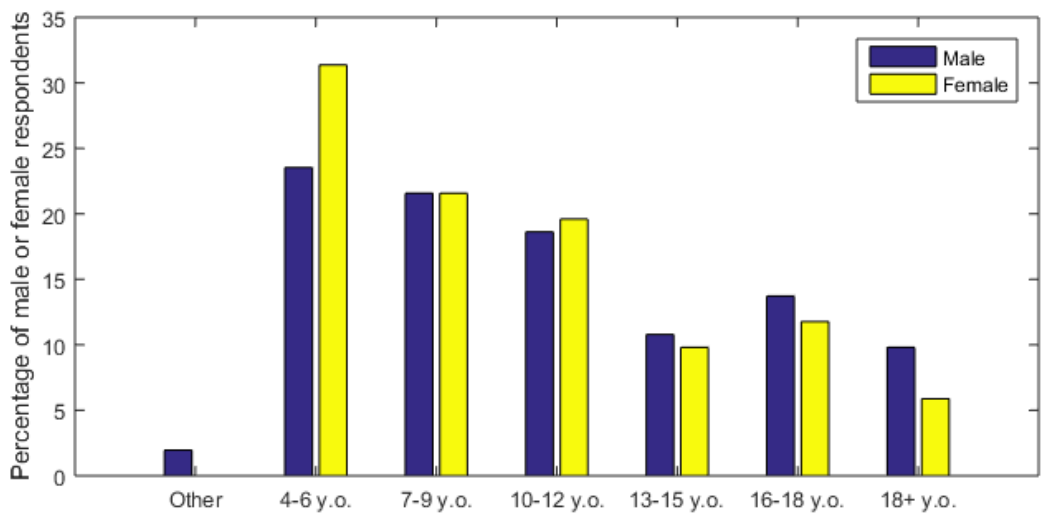

*Figure 1 Age of initial interest in astronomy.*

Nevertheless, the decision to become a professional astronomer or to pursue studies to become an astronomer happened later. Over half of the participants claimed this motivation occurred during the end of secondary school or undergraduate studies. For many, taking an introductory course to astronomy during their undergraduate studies at universities was the decisive factor. Similarly to Lindahl (2007), even though the first interest for science occurs at a fairly young age, if their interaction with science during high school or freshman years of university is not engaging, this can dissuade them from choosing a career in science. This implies the decision about career path occurs later than childhood or early adolescence.

But what ignited the astronomers' interest in the subject? Over a quarter of our participants did not have a specific starting point, but those who did pointed to their inspiration from the night sky (42%) or their excitement after reading a popular science book (32%) . Additionally, looking through a telescope and following the Apollo Mission appears to be of great importance initiating the first spark for astronomy along with being inspired from a family member or a teacher. This shows that first interest in science often happens outside of classroom, which is also in agreement with Lindahl (2007) which reported that participants did not find science taught in class particularly engaging nor representative of the natural world they experience in everyday life. These results draw out the importance of science education and public outreach among children at primary schools and ensure that teenagers maintain an enjoyable experience with science.

## 4. Views on Education and Public Outreach

Many studies revealed that most scientists have a positive attitude towards public outreach initiatives (Poliakoff, 2007; Andrews, 2005; Ecklund, 2010). Our study showed similar results as 79% of the respondents think that education and public outreach initiatives are essential and 19% claim it is important. Another way of assessing the views of education and public outreach initiatives is to evaluate the amount of time and financial support scientists dedicate to EPO. For completeness reasons, participants were given the possibility to not disclose answers questions concerning the budget and time spent on outreach initiatives.

As studies showed many scientists viewed EPO as a hobby rather than as part of their duty at work, the participants were asked for the amount of free time and working time spent on EPO activities (Poliakoff, 2007). The analysis of the data revealed no significant difference between the amount of free time and working time allocated for EPO activities with a median of 0 to 2 hours spent on EPO per week on average as shown in Figure 2. After using the Spearman's rho correlation, it was determined that the scientists who claim to spend more time at work on EPO activities weekly also dedicate more time outside of work. The analysis showed a moderate correlation between the two variables ($\rho$= 0.46; $p<0.05$). Interestingly, this does not agree with Poliakoff's study which reported that scientists considered EPO activities to be a hobby rather than a work duty. This implies that time constraint isn't the main factor influencing astronomers to take part of outreach activities, and that there exist other factors motivating them into investing both time at work and outside of work to such projects.

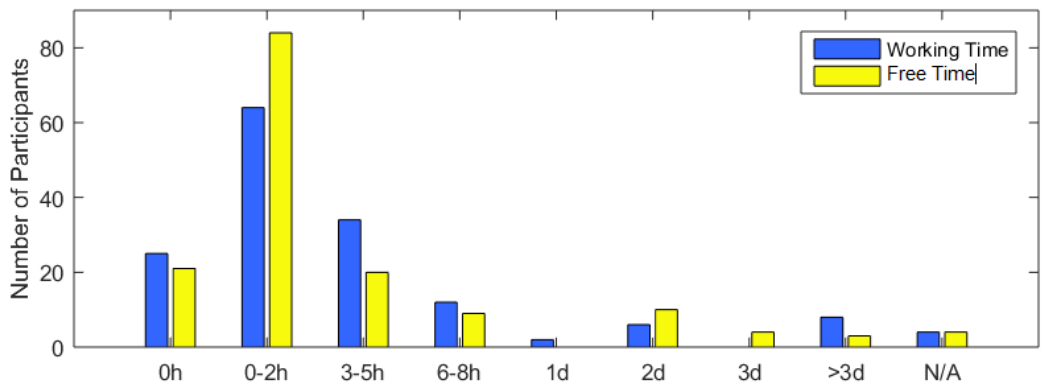

*Figure 2 Distribution of working and free time spent on average on EPO activities per week.*

Out of the 155 respondents, a quarter of them (56 participants) chose to not disclose the percentage of their research grant attributed to EPO. Among those who did answer the question (N=116), 50 astronomers claimed that 0% of their grant money was allocated to education and public outreach and 15 of them use between 0-2% for EPO activities. Hence, most the respondents reported that less than 2% of their research grant into EPO initiatives, which is less financial support than what is suggested in many science communication guidelines (Brake, 2010; Bowater, 2013). As mentioned before, the 2000 Wellcome Trust report showed there was a lack of financial support for EPO.

To explore this matter, astronomers were asked what percentage of grant money should be invested into EPO. The response rate for this question was 83% (138 out of 155 participants). Interestingly, the results significantly differed from the previous question ($p < 0.05$) as shown in Figure 3. This time, only 13 respondents claimed that 0% of research grants should be invested into EPO activities. On average astronomers suggested that 5-10% of research grant should be allocated to EPO activities which is significantly greater than the amount used for outreach.

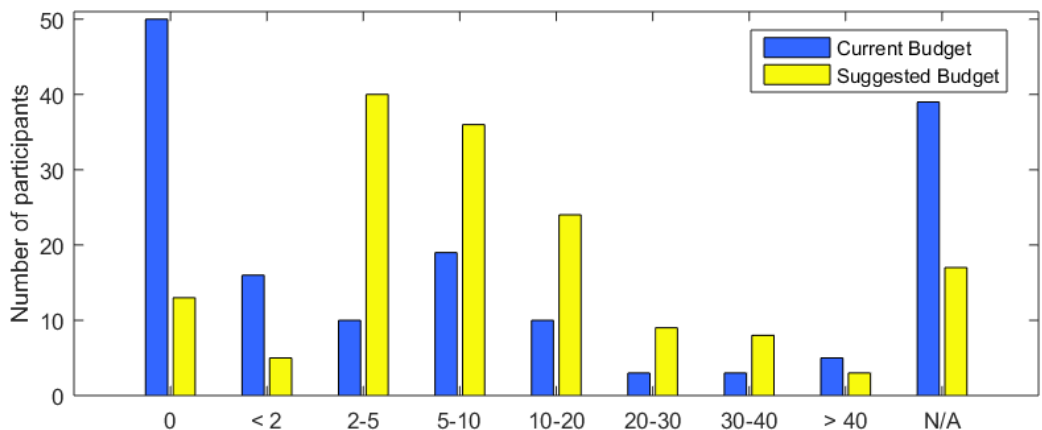

*Figure 3 Distribution of percentage of research grant astronomers currently invest and suggest to allocate into public outreach engagement.*

Given this result, a new question arises, do astronomers generally wish to spend more of their research grant into EPO than what they currently spend? The Spearman rank correlation test revealed a correlation between respondents' current and suggested budget spent on EPO activities ($\rho = 0.59; p < 0.05$). This shows that in general in this survey, the participants suggested a higher amount of portion of their research grant than what is currently allocated to outreach initiatives. This implies that astronomers generally think there is a lack of financial support for EPO activities and suggest that policies on the distribution of their research grant includes a higher budget for EPO.

An interesting finding from Poliakoff's studies on factors predicting scientists' decision to participating in EPO activities was their past behavior. The research revealed that a scientist who has been involved in EPO projects in the past is more likely to participate in the near future (upcoming year). Consequently, taking part of outreach activities at an early stage of career increases the chance that a scientist would get involved in EPO activities regularly in future career stages. However, Ecklund's studies on views of public engagement activities among scientists demonstrate that one of the participants' concern was the lack of support from mentors for taking part in outreach activities. As a result, this also affects scientists' decision to take part in of outreach projects at later stages of their careers. To address this topic, astronomers were asked if they recommended or encouraged their student to get involved with EPO projects. For the most part (70% of the participants), the answer was positive as opposed to 2% who answered negatively. The majority of the 43 participants who did not answer the question were either Master/PhD students or post-doctoral fellows for whom the question was not applicable. This was unexpected since many scientists claimed a factor inhibiting the participation in EPO initiatives was the disapproval by mentors and department heads (Ecklund, 2012). This could either mean that encouragement to participate in outreach projects is more present in the community of astronomers than other sciences. However, the way the question was posed was biased towards a positive response. A more accurate way to measure this would have been give the respondents an ordinal scale rather than the only possibilities of answering positively or negatively when they were asked if they encourage their students to participate in EPO initiatives.

## 5.    Conclusion

Astronomers' interest in astronomy is shaped from a very young age (between 4 to 6 years old).As expected, despite the gender imbalance in the astronomical field, there is no evidence of a difference in the age of development of interest for astronomy between genders. Furthermore, the decision of undertaking a career in astronomy only comes later during the end of adolescence and early adulthood, which shows the importance of nurturing their astronomy up to this age.

Most astronomers claim to have a positive attitude towards education and public outreach as the majority think it is valuable and those in authority encourage their students to participate in outreach. However, other studies have shown that although scientists have a positive view of EPO, some of them do not participate because of their mentors and department heads disapproval which is inconsistent with our results. The results showed that astronomers allocated less time into EPO on average than what was recommended by practical guides for science communication. Some science communication books explain this with the view scientists have about EPO: science communication projects are initiatives they would only undertake if they have extra time beside their other duties in academia. The study also disproved the theory that education and public outreach activities were viewed as a hobby rather than a duty as there was no significant difference between the time astronomers put into EPO during working time and free time. Interestingly, the analysis showed that those who claimed to spend more working time on EPO activities also invested as much of their free time. This implies that there must be some factors other than the views on their responsibility as a scientist which motivates working on EPO project both at work and outside of work.

Another finding is that the percentage of research grant astronomers allocate to EPO does not align with the amount they suggested. In general, most of them suggested a higher amount than what they currently attributed to EPO activities. This indicates that astronomers would like to invest a larger ratio of their grant towards EPO and therefore, calls for a change in grant policies.

## 6.    Limitations and Future Work

As the collection of data did not keep track of the number of solicited astronomers who chose not to participate to the survey, this prevented us from obtaining a response rate to evaluate if the sample was a good representation of the target group. Second of all, on-line surveys also imply a certain level of self-selection bias; which is a limitation of this study. Some questions about the value of EPO and the encouragement in taking part of EPO activities were asked in a manner biased towards a positive answer. The data obtained revealed some interesting insights but it is not complete. A more detailed study about the views of astronomers would allow us to tackle down the motivating and inhibiting factors for public outreach initiatives. It will also give more insight about the changes that need to be made in order to improve the number of astronomers participating in EPO and how these changes can be implemented. The next goal is to explore astronomers' points of view on different aspects of their attitude towards engagement initiatives, their level of confidence in taking part in EPO activities, their perception of the public and their peers and the value of education and public outreach at their institution.

## 7.   Data

The source data files used in this paper are available on an open repository: github.com/unawe/research/

## 8.   Acknowledgments

We would like to thank Valério A. R. M. Ribeiro, Thilina Heenatigala, Avivah Yamani, Mara van Beusekom, Gimenne Zwamam, Joshua Borrow and Jack Sankey for their support and contribution for this article.

## 9.   References

Andrews, E., Weaver, A., Shamatha, J., Melton, G., & Hanley, D., 2005, *Scientists and Public Outreach: Participation, Motivations, and Impediments*, Journal of Geoscience Education, 281

Bowater, L., & Yeoman, K., 2013, *Science communication*, (Hoboken: Wiley)

Brake, M., & Weitkamp, E., 2010, *Introducing science communication*, (Houndmills, Basingstoke, Hampshire: Palgrave Macmillan)

Bulunuz, M., & Jarrett, O., 2010, *Developing an interest in science: background experiences of preservice elementary teachers*, International Journal of Environmental & Science Education, *5*, 65

Bush, G. W, 2006, *State of the Union address (Copy of text from speech)*, http://www.whitehouse.gov/stateoftheunion/2006/, June 2015

Burns, T., O'Connor, D., & Stocklmayer, S, 2003, *Science Communication: A Contemporary Definition*, Public Understanding Of Science, *12*, 183

Christensen, L., 2007, *The hands-on guide for science communicators*. (New York: Springer)

Cleaves, A.,2005, *The formation of science choices in secondary school,* International Journal of Science Education, 27, 471

Ecklund, E., James, S., & Lincoln, A., 2012, *How Academic Biologists and Physicists View Science Outreach,* Plos ONE, *7*, e36240

European Commission, 2004, *Report by the High Level Group on increasing human resources for science and technology in Europe*, (Luxembourg: Office for Official Publications of the European Communities)

Jensen, P., Rouquier, J., Kreimer, P., & Croissant, Y., 2008, *Scientists Connected with the Society are more Active Academically*, Science and Public Policy, 35, 527

Lindahl, B.,2007, *A longitudinal study of students' attitudes towards science and choice of career*, (Paper presented at annual meeting of the National Association for Research in Science Teaching, New Orleans, LA)

Maltese A. & Tai R., 2010, *Eyeballs in the Fridge: Sources of early interest in science*, International Journal of Science Education, 32, 669

Mathews, D., Kalfoglou, A., & Hudson, K., 2005, *Geneticists' views on science policy formation and public outreach*, American Journal Of Medical Genetics Part A, *137A*, 161

Poliakoff, E., & Webb, T., 2007, *What Factors Predict Scientists' Intentions to Participate in Public Engagement of Science Activities?*, Science Communication, *29*, 242

Rosenberg, M., Russo, P., Bladon, G., & Christensen, L., 2013, *International Astronomical Union | IAU. Iau.org*, http://www.iau.org/public/themes/why_is_astronomy_important/, June 2015

Royal Society, 2004, *Taking a leading role: A good practice guide (Scientist survey)*, http://www.royalsoc.ac.uk/page.asp?id=2903, June 2015

Wellcome Trust, 2000, *Science and the Public: Public Attitudes to Science in Britain.* (London: The Wellcome Trust Publishing Department)

Wellcome Trust, 2000, *The role of scientists in public debate*, (London: The Wellcome Trust Publishing Department)

# 10. Appendix A: Survey

## Astronomers' Views on Education and Public Outreach

## Questionnaire

**Personal data**
Name (optional): _________________________________
Institution: _______________________________
Country: _________________________________
Age: _____________
Gender:  M  F  N/A

Academic level:

- o  Undergrad Student
- o  Master Student
- o  PhD Student
- o  Post Doc
- o  Faculty
- o  Director/Head
- o  Other, specify: _____________

**Interest in Science/Astronomy**
At what age did you develop an interest in science/astronomy?

- o  4-6 years
- o  7-9 years
- o  10-12 years
- o  13-15 years
- o  16-18 years
- o  18+ years
- o  Other, specify: _____________

Did you have a specify moment that sparked your interest in astronomy?

- o  Yes
- o  No

If yes, what sparked your  interest?

- o  Reading a popular science book (specify: __________________)
- o  Watching a TV series/documentary
    - o  (specify: __________________)
- o  Looking through a telescope
- o  Becoming a member of an astronomical society

- Inspired by the night sky
- Inspired by your parent
- Inspired by a teacher
    - (which subject?___________________)
- Other, specify: ____________________

When did you decide to become an astronomer?

- 4-6 years
- 7-9 years
- 10-12 years
- 13-15 years
- 16-18 years
- 18+ years
- Other, specify: ________________

## Views on Education and Public Outreach(EPO)

How valuable do you consider educational and public outreach initiatives?

- Essential
- Important
- Neutral
- Not important
- Useless

What percentage of your current research budget is spent on EPO activities?

____________ %

In your opinion, what percentage of a research grant **should** be used for EPO activities on average?

_____________ %

How much of your working time do you spend on EPO activities?

- 0
- <2 hours/week
- 3-5 hours/week
- 6-8 hours/week
- 2 days/week
- 3 days/week
- > 3 days/week
    - Specify: _____________

How much of your own free time do you spend on EPO activities?

- 0

- < 2 hours/**month**
- 3-5 hours/**month**
- 6-8 hours/**month**
- 2 days/**month**
- 3 days/**month**
- >3 days/**month**
  - Specify: _______________

Do you recommend/encourage your students/Post-docs to be involved in EPO?

- Yes
- No

How do you participate in EPO activities?

- Presenting at schools
- Bringing students into the lab/department
- Giving public lectures
- Writing popular articles
- Writing popular books
- Writing/contributing to a blog
- Supporting EPO activities in a planetarium
- Observing events
- Institution/Observatories Open days
- Supporting exhibitions development
- Other: _______________
- I don't participate

Please write down your email address here if you would like to be informed about IAU EPO activities and the results of this questionnaire: _________________________________________